\documentclass{article}
\usepackage{amsmath}
\usepackage{graphicx}

\begin{document}

\title{Superluminal phase and group velocities: A tutorial on Sommerfeld's
phase, group, and front velocities for wave motion in a medium, with
applications to the \textquotedblleft instantaneous
superluminality\textquotedblright\ of electrons.}
\date{November 9, 2011}
\author{Raymond Chiao (rchiao@ucmerced.edu) \\
University of California at Merced, P. O. Box 2039, Merced, CA 95344}
\maketitle

\section{Introduction}

In 1905, Einstein published his historic paper on special relativity.
Shortly afterwards, Sommerfeld \cite{Brillouin}\ answered criticisms of
Einstein's work, viz., that the phase and group velocities of
electromagnetic waves can become superluminal, since these two kinds of
velocities can exceed the vacuum speed of light\ inside a dielectric medium.
Note that Einstein considered wave propagation solely \emph{in the vacuum},
whereas his critics considered wave propagation \emph{in media}.

Sommerfeld pointed out that while it is true that both the phase and the
group velocities in media can in fact exceed $c$, the front velocity,
defined as the velocity of a \emph{discontinuous jump} in the initial wave
amplitude from zero to a finite value, cannot exceed $c$. It is Sommerfeld's
principle of the \emph{non-superluminality} \emph{of the front velocity}
that prevents a violation of the Einstein's basic principle of causality in
special relativity, i.e., that no effect can ever precede its cause.

In subsequent work, Sommerfeld and Brillouin \cite{Brillouin} showed that
the \textquotedblleft front\textquotedblright\ is accompanied by two kinds
of \textquotedblleft precursors\textquotedblright , now known as the
\textquotedblleft Sommerfeld\textquotedblright , or the \textquotedblleft
high-frequency\textquotedblright , precursor, and the \textquotedblleft
Brillouin\textquotedblright , or the \textquotedblleft
low-frequency\textquotedblright , precursor. These precursors are weak
ringing waveforms that \emph{follow} the abrupt onset of the front, but they 
\emph{precede} the gradual onset of the strong main signal.

\section{Superluminal phase velocities}

It is well known that the phase velocity of electromagnetic waves can become
superluminal under certain circumstances. A simple example is the
superluminality of the phase velocity of an electromagnetic wave traveling
within a rectangular waveguide in its fundamental TE$_{01}$ mode. We shall
give below yet another, more impressive, example, namely, the
superluminality of the phase velocity of X-rays in all materials. As was
first noticed by Einstein, the superluminality of the phase velocity of
X-rays in all kinds of crystals leads to the phenomenon of total \emph{%
external} reflection of X-rays impinging at grazing incidence from the
vacuum upon the surfaces of these crystals.

The definition of the phase velocity is best given through an example.
Consider a monochromatic electromagnetic plane wave traveling down the $z$
axis of a homogeneous dielectric medium%
\begin{equation}
E(z,t)=E_{0}\cos (kz-\omega t)  \label{E field of plane wave}
\end{equation}%
\begin{equation}
B(z,t)=B_{0}\cos (kz-\omega t)
\end{equation}%
where $\omega $ is the angular frequency of the wave and $k$ is its
wavenumber. The phasefronts $\phi (z,t)$ of this wave are defined through
the relationship%
\begin{equation}
\phi (z,t)=kz-\omega t=k(z-\frac{\omega }{k}t)=k(z-v_{\rm{phase}}t)=\rm{%
const}
\end{equation}%
where the phase velocity $v_{\rm{phase}}$ is defined as follows:%
\begin{equation}
v_{\rm{phase}}=\frac{\omega }{k}
\end{equation}%
Thus a given phasefront of the electromagnetic wave satisfies the
relationship%
\begin{equation}
z-v_{\rm{phase}}t=\frac{\rm{const}}{k}=\rm{const}^{\prime }=z_{0}
\end{equation}

It is customary to define the index of refraction $n\left( \omega \right) $
of the medium through the relationship%
\begin{equation}
k=k\left( \omega \right) =n\left( \omega \right) \frac{\omega }{c}
\end{equation}%
where $c$ is the vacuum speed of light. Thus the phase velocity is related
to the index of refraction by%
\begin{equation}
v_{\rm{phase}}=\frac{\omega }{k\left( \omega \right) }=\frac{c}{n\left(
\omega \right) }
\end{equation}%
For a typical transparent medium, such as a piece of glass, the index of
refraction is greater than unity, so that the phase velocity of light in
glass is less than the vacuum speed of light.

However, the index of refraction function $n\left( \omega \right) $ is in
general determined by the dispersive properties of the medium, and can be
less than unity. For example, consider a medium consisting of Lorentz
oscillators which obey the simple harmonic equation of motion%
\begin{equation}
\ddot{x}+\gamma \dot{x}+\omega _{0}^{2}x=eE/m  \label{SHO EOM}
\end{equation}%
where $\gamma $ is the damping constant of the oscillator, $\omega _{0}$ is
its resonance frequency, $e$ is the charge of an electron, and $m$ is its
mass. The solution of the equation of motion (\ref{SHO EOM}) of the simple
harmonic oscillator, when it is being driven by the monochromatic electric
field $E$ written in its complex exponential form,%
\begin{equation}
E=E_{0}\exp \left( ikz-i\omega t\right) 
\end{equation}%
is given by%
\begin{equation}
x=\frac{eE/m}{\omega _{0}^{2}-\omega ^{2}-i\gamma \omega }
\end{equation}%
The polarization $P$ of the medium is therefore given by%
\begin{equation}
P=n_{\rm{atoms}}ex=\frac{n_{\rm{atoms}}e^{2}E/m}{\omega _{0}^{2}-\omega
^{2}-i\gamma \omega }=\varepsilon _{0}\chi E
\end{equation}%
when $n_{\rm{atoms}}$ is the number density of atoms, that is, the number
density of Lorentz oscillators. Solving for the susceptibility of the medium 
$\chi $, one then finds that%
\begin{equation}
\chi =\frac{n_{\rm{atoms}}e^{2}/m\varepsilon _{0}}{\omega _{0}^{2}-\omega
^{2}-i\gamma \omega }=\frac{\omega _{p}^{2}}{\omega _{0}^{2}-\omega
^{2}-i\gamma \omega }
\end{equation}%
where the plasma frequency is defined as%
\begin{equation}
\omega _{p}=\left( n_{\rm{atoms}}e^{2}/m\varepsilon _{0}\right) ^{1/2}
\end{equation}%
The dielectric constant of the medium is related to the susceptibility by
the definition%
\begin{equation}
\varepsilon =1+\chi 
\end{equation}%
Therefore the index of refraction for a medium of Lorentz oscillators is given by%
\begin{equation}
n=\sqrt{\varepsilon }=\sqrt{1+\chi }=\sqrt{1+\frac{\omega _{p}^{2}}{\omega
_{0}^{2}-\omega ^{2}-i\gamma \omega }}  \label{index for Lorentz model}
\end{equation}

Einstein noticed that for sufficiently high frequency X-rays, i.e., those
X-rays which have a frequency above the highest possible X-ray transition
frequency of the atoms in a crystal, namely, those transitions in which the
most tightly bound electron in the ground state of the atom (i.e, the
electron in the 1S state, or K shell, closest to the nucleus) is knocked out
by the X-ray into the continuum, one can approximate the Lorentz model for
the index of refraction (\ref{index for Lorentz model}) by its
high-frequency form%
\begin{equation}
n_{\rm{X-ray}}=n\left( \omega \rightarrow \infty \right) =\sqrt{1+\frac{%
\omega _{p}^{2}}{\omega _{0}^{2}-\omega ^{2}-i\gamma \omega }}\rightarrow 1-%
\frac{1}{2}\frac{\omega _{p}^{2}}{\omega ^{2}}<1
\label{phase velocity for X-rays}
\end{equation}
\emph{In other words, the phase velocity of sufficiently high-frequency
X-rays in any medium that can be modeled by Lorentz oscillators will always
be superluminal. }Thus it follows that the index of refraction of any kind
of crystal for X-rays of sufficiently high energy will always be slightly
less than unity. However, from (\ref{phase velocity for X-rays}), it can be
shown that the group velocity for these same X-rays in the same crystal will
always be subluminal, since the group index (see (\ref{group index})),
which, at sufficiently high frequencies, is given by%
\begin{equation}
n_{\rm{X-ray}}+\left. \omega \frac{dn}{d\omega }\right\vert _{\rm{X-ray}%
}\rightarrow 1+\frac{1}{2}\frac{\omega _{p}^{2}}{\omega ^{2}}>1
\label{group velocity for X-rays}
\end{equation}%
will always exceed unity.

\begin{figure}
\includegraphics[width=4.75in]{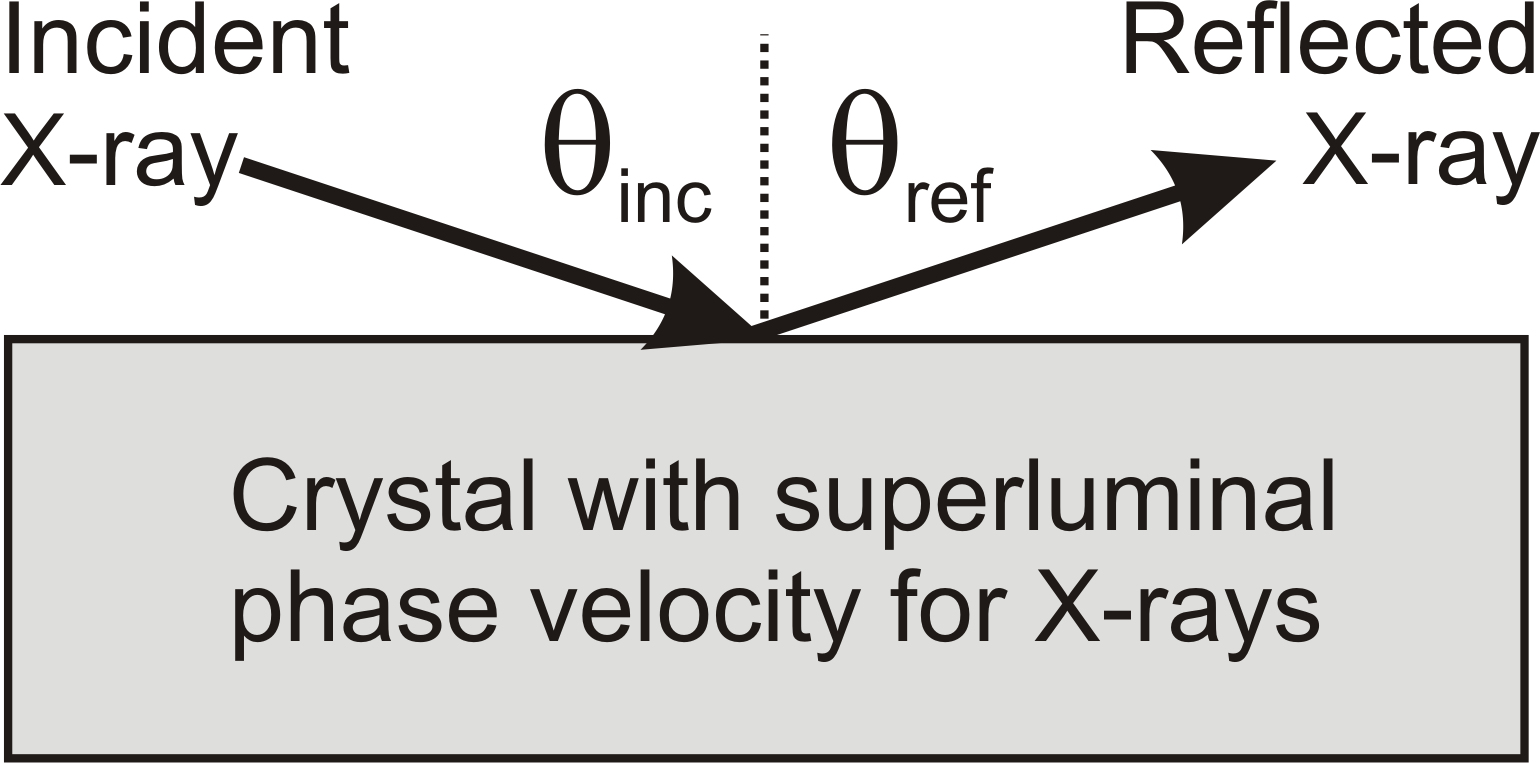}
\caption{Einstein's total \emph{external}
reflection of grazing X-rays from a crystal of Lorentz oscillators. The
phase velocity for sufficiently energetic X-rays inside the crystal will
always be superluminal.}
\end{figure}

Next, Einstein pointed out that Snell's law, when applied to the
vacuum-medium interface in Figure 1 using (\ref{phase velocity for X-rays}),
will lead to a critical angle given by%
\begin{equation}
\sin \theta _{\rm{crit}}=n_{\rm{X-ray}}\cdot \sin 90^{\circ }=n_{\rm{%
X-ray}}<1
\end{equation}%
Since the index of refraction is less than unity, there always will exist a
solution for the critical angle for the total \emph{external} reflection of
grazing-incidence X-rays%
\begin{equation}
\theta _{\rm{crit}}=\sin ^{-1}n_{\rm{X-ray}}
\end{equation}%
The complement of this critical angle, $\phi _{\rm{crit}}=90^{\circ
}-\theta _{\rm{crit}}$, which, for grazing incidence X-rays, is given by%
\begin{equation}
\sin \theta _{\rm{crit}}=\cos \phi _{\rm{crit}}\approx 1-\frac{1}{2}\phi
_{\rm{crit}}^{2}\approx 1-\frac{1}{2}\frac{\omega _{p}^{2}}{\omega ^{2}}
\end{equation}%
so that the complement of the critical angle is given approximately by%
\begin{equation}
\phi _{\rm{crit}}\approx \frac{\omega _{p}}{\omega }<<1
\label{critical phi}
\end{equation}%
Now the plasma frequency lies in the ultraviolet part of the electromagnetic
spectrum, whereas X-rays lie at much higher frequencies than ultraviolet
frequencies. For example, for a 1 keV X-ray reflecting from gold, whose
plasma frequency corresponds to an energy of approximately 30 eV \cite{Pines}%
, $\phi _{\rm{crit}}$ is known to be 3.72 degrees \cite{X-ray telescopes},
which agrees with (\ref{critical phi}) within a factor of two. 

Satellite X-ray telescopes use grazing-incidence optics, which is based on
Einstein's total-external-reflection effect, in order to form X-ray images
of distant astrophysical objects \cite{X-ray telescopes}. This demonstrates
that superluminal phase velocities have important applications.

It is a common misconception that superluminal phase velocities are
unobservable \cite{Commins}. However, as the above example of Einstein's
total external reflection of X-rays shows, there exists at least one
striking counter-example that can serve to dispel this misconception.

\section{Superluminal signaling is not possible using superluminal phase
velocities}

Can one send a true signal faster than light by means of a superluminal
phase velocity? The answer is no, since the phase velocity is the velocity
of the crests (i.e., the phasefronts) of a continuous-wave, monochromatic,
electromagnetic wave. Since the amplitude and phase of a continuous wave is
not changing with time, there can be no information contained within such a
waveform. As in radio, one must introduce a truly \emph{time-dependent
modulation} of a continuous \textquotedblleft carrier\textquotedblright\
waveform (using either AM or FM modulation), before any true signal can be
sent via the carrier wave.

In the case of quantum mechanics, the wavefunction of an electron can be
written in terms of an amplitude and a phase factor as follows:%
\begin{equation}
\psi \left( \mathbf{r},t\right) =A\left( \mathbf{r},t\right) e^{i\phi \left( 
\mathbf{r},t\right) }
\end{equation}%
For the special case of a monochromatic plane wave traveling in the $z$
direction%
\begin{equation}
A\left( z,t\right) =A_{0}
\end{equation}%
is a constant, and
\begin{equation}
\phi \left( z,t\right) =kz-\omega t
\end{equation}%
so that the electron wavefunction in a momentum eigenstate has the form%
\begin{equation}
\psi \left( z,t\right) =A_{0}e^{i\left( kz-\omega t\right) }
\end{equation}%
Therefore the phase velocity of the electron is determined through the
relationship%
\begin{equation}
\psi \left( z,t\right) =A_{0}e^{i\left( kz-\omega t\right)
}=A_{0}e^{ik\left( z-\frac{\omega }{k}t\right) }=A_{0}e^{ik\left( z-v_{\rm{%
phase}}t\right) }
\end{equation}%
so that once again%
\begin{equation}
v_{\rm{phase}}=\frac{\omega }{k}
\end{equation}

Now the Born interpretation tells us that the probability density for
finding the electron at $\left( z,t\right) $ is given by%
\begin{equation}
\left\vert \psi \left( z,t\right) \right\vert ^{2}=\left\vert
A_{0}e^{i\left( kz-\omega t\right) }\right\vert ^{2}=\left\vert
A_{0}\right\vert ^{2}=\rm{constant}  \label{abs sq of wf}
\end{equation}%
and is therefore constant for a plane wave. Since the overall phase factor
of the wavefunction is not an observable quantity, it can be argued that the
phase velocity of the electron is not an observable quantity. However, the
overall phase factor picked up by the wavefunction can in fact be observed
in interference experiments.

However, in the case of a classical electromagnetic plane wave, the phase
velocity \emph{is} an observable quantity, just as the speed of the crests
of the ripples of water waves on the surface of a pond is obviously an
observable quantity. Furthermore, the Poynting vector for a classical
electromagnetic plane wave is given by%
\begin{equation}
\mathbf{S}=\mathbf{E\times H}=\mathbf{\hat{k}}\left( E_{0}H_{0}\right) \cos
^{2}(kz-\omega t)=\mathbf{\hat{k}}\left( E_{0}H_{0}\right) \left( \frac{1}{2}%
-\frac{1}{2}\cos (2kz-2\omega t)\right)
\end{equation}%
which clearly has a second-harmonic component that moves at the phase
velocity $\omega /k$, which can in principle be observed. No such
second-harmonic component exists in the case of the electron, as is evident
by inspection of (\ref{abs sq of wf}). However, although the phase velocity
of an electromagnetic plane wave is in principle an observable quantity, no
true signal can be transmitted by means of it.

\section{Superluminal group velocities}

While it is well known that phase velocities can become superluminal, it is
less well known that group velocities can also become superluminal. There is
a common misconception that the group velocity is the \textquotedblleft
signal\textquotedblright\ velocity of physics, which relates a cause to its
effect. However, as we shall presently see, the group velocity is \textit{not%
} the velocity that relates a cause to its effect. Only the front velocity
can fulfill this role.

Consider a wavepacket propagating along the $z$ axis. In quantum theory,
such a wavepacket, for example a Gaussian wavepacket containing a single
electron within it, can be represented by the Fourier integral%
\begin{equation}
\psi (z,t)=\int\limits_{-\infty }^{\infty }d\omega \tilde{\psi}\left(
\omega \right) e^{ik(\omega )z-i\omega t}  \label{FT wavepacket}
\end{equation}%
Suppose that the wavepacket is strongly peaked in its amplitude $\tilde{\psi}%
\left( \omega \right) $ at some frequency $\omega _{0}$. It is natural then
to perform a Taylor series expansion of the wavenumber $k(\omega )$ around $%
\omega _{0}$, which yields%
\begin{equation}
k\left( \omega \right) =k\left( \omega _{0}\right) +\left( \omega -\omega
_{0}\right) \left. \frac{dk}{d\omega }\right\vert _{\omega _{0}}+...
\end{equation}%
One can therefore approximate the Fourier integral (\ref{FT wavepacket}) as
follows:%
\begin{eqnarray}
\psi (z,t) &=&\int\limits_{-\infty }^{\infty }d\omega \tilde{\psi}\left(
\omega \right) \exp \left( ik(\omega _{0})z+i\left( \omega -\omega
_{0}\right) \left. \frac{dk}{d\omega }\right\vert _{\omega
_{0}}z+...-i\omega t\right)   \nonumber \\
&=&\int\limits_{-\infty }^{\infty }d\omega \tilde{\psi}\left( \omega
\right) e^{ik(\omega _{0})z+i\left( \omega -\omega _{0}\right) \left. \frac{%
dk}{d\omega }\right\vert _{\omega _{0}}z+...-i\left( \omega -\omega
_{0}\right) t}e^{-i\omega _{0}t}  \nonumber \\
&=&e^{ik(\omega _{0})z-i\omega _{0}t}\int\limits_{-\infty }^{\infty
}d\omega \tilde{\psi}\left( \omega \right) e^{i\left( \omega -\omega
_{0}\right) \left. \frac{dk}{d\omega }\right\vert _{\omega _{0}}z-i\left(
\omega -\omega _{0}\right) t+...}  \nonumber \\
&=&e^{ik(\omega _{0})z-i\omega _{0}t}\int\limits_{-\infty }^{\infty
}d\omega \tilde{\psi}\left( \omega \right) e^{i\left( \omega -\omega
_{0}\right) \left. \frac{dk}{d\omega }\right\vert _{\omega _{0}}\left(
z-\left. \frac{dk}{d\omega }\right\vert _{\omega _{0}}^{-1}t\right) +...} 
\nonumber \\
&\propto &\psi (z-v_{\rm{group}}t)
\end{eqnarray}%
where the group velocity is identified as%
\begin{equation}
v_{\rm{group}}=\left. \frac{dk}{d\omega }\right\vert _{\omega
_{0}}^{-1}=\left. \frac{d\omega }{dk}\right\vert _{\omega _{0}}
\end{equation}%
The meaning of the group velocity is that it is the velocity with which the
peak of the wavepacket moves.

Now for an optical medium, we saw earlier that%
\begin{equation}
k\left( \omega \right) =\frac{n\left( \omega \right) \omega }{c}
\end{equation}%
where $n\left( \omega \right) $ is medium's refractive index. It follows that%
\begin{equation}
\frac{dk\left( \omega \right) }{d\omega }=\frac{1}{c}\left( n\left( \omega
\right) +\omega \frac{dn\left( \omega \right) }{d\omega }\right) 
\end{equation}%
and therefore that the group velocity is given by%
\begin{equation}
v_{\rm{group}}=\left. \frac{dk\left( \omega \right) }{d\omega }\right\vert
_{\omega _{0}}^{-1}=\frac{c}{\left. n\left( \omega \right) +\omega \frac{%
dn\left( \omega \right) }{d\omega }\right\vert _{\omega _{0}}}
\label{group velocity with denominator}
\end{equation}%
The denominator of this expression for the group velocity%
\begin{equation}
n_{\rm{group}}=\left. n\left( \omega \right) +\omega \frac{dn\left( \omega
\right) }{d\omega }\right\vert _{\omega _{0}}  \label{group index}
\end{equation}%
is called the \textquotedblleft group index\textquotedblright . By
inspection of the group index, it is apparent that it can vanish whenever%
\begin{equation}
n\left( \omega \right) +\omega \frac{dn\left( \omega \right) }{d\omega }=0
\label{infinite group velocity}
\end{equation}%
This can happen whenever there is anomalous dispersion%
\begin{equation}
\frac{dn\left( \omega \right) }{d\omega }\,<0
\end{equation}%
i.e., whenever the index of refraction decreases with increasing frequency.
Whenever this can happen, the group velocity can become infinite, which is
obviously a kind of superluminal behavior.

\begin{figure}
\includegraphics[width=4.75in]{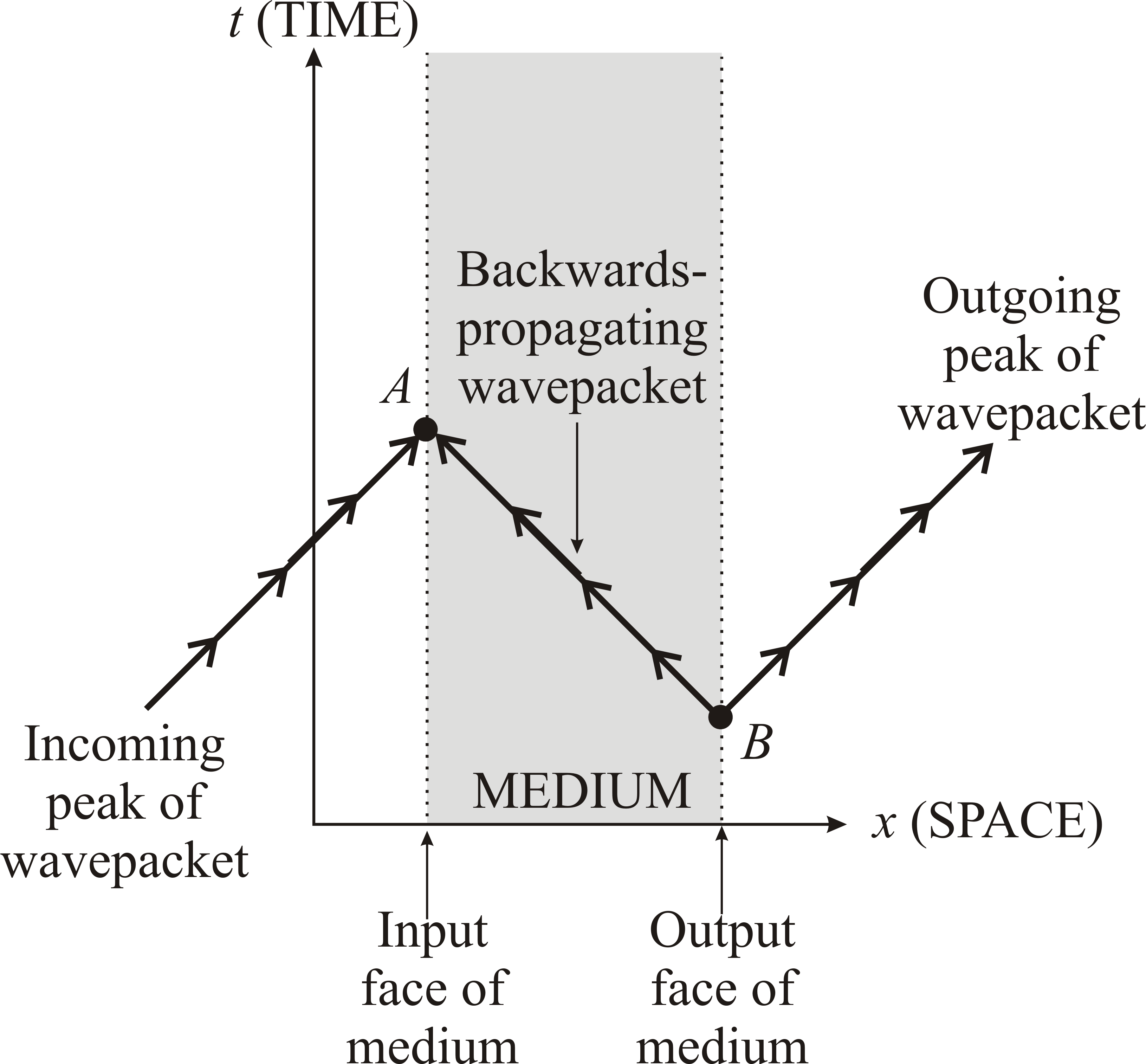}
\caption{Feynman-like space-time diagram
for the motion of the peaks of a wavepacket propagating with a negative
group velocity inside a superluminal medium (in \emph{gray}).}
\end{figure}

By inspection of the denominator of (\ref{group velocity with denominator}),
it is also clear that the group velocity can become \emph{negative} whenever
the group index is negative, i.e., whenever%
\begin{equation}
n\left( \omega \right) +\omega \frac{dn\left( \omega \right) }{d\omega }\,<0
\end{equation}%
The meaning of a negative group velocity is this: \emph{Before} the peak of
an incoming wavepacket has entered the entrance face of the medium, the peak
of an outgoing wavepacket has \emph{already} left the exit face of medium.
This highly counter-intuitive, superluminal behavior in fact does not
violate causality, and has in fact been observed in many experiments \cite%
{Shanghai superluminality}. It can be understood with the help of the
Feynman-like space-time diagram in Figure 2.

By taking an arbitrary time-slice between the two events $A$ and $B$ in this
space-time diagram, one sees that there exist \emph{three} wavepackets at
this moment of time. The first wavepacket is the one coming\ in from the
left towards the input face of the medium, the second wavepacket is the one
propagating backwards within the medium from the output face of the medium
towards the input face of the medium, and the third wavepacket is the one
leaving the output face of the medium, and going out towards the right.
Event $A$ corresponds to an event in which the first and second wavepackets
annihilate with each other in a \textquotedblleft pair
annihilation\textquotedblright\ event at the input face, and event $B$
corresponds to an event in which the second and third wavepackets are
created together in a \textquotedblleft pair creation\textquotedblright\
event at the output face. Note that the pair-creation event precedes the
pair-annihilation event.

In computer simulations of this phenomenon, it is observed that the early,
analytic tail of a Gaussian wavepacket penetrates deeply into the medium.
This early tail of the incoming wavepacket then triggers the emission of the
pair of wavepackets at the pair-creation event $B$ at the output face of the
medium. The backwards-propagating wavepacket is then observed to
subsequently annihilate with the incoming wavepacket at the
pair-annihilation event $A$ at the input face of the medium. In the special
case of an inverted two-level medium excited far off of resonance by the
incident wavepacket, the medium loans energy from the inverted two-level
system in order to produce the two new wavepackets at $B$. This the loan is
repaid later to the medium at $A$.

Not only can the group velocity become negative, but under certain
circumstances it also can become infinite, such as in the special case (\ref%
{infinite group velocity}). We shall call this important special case
\textquotedblleft instantaneous superluminality\textquotedblright , and we
shall see that it can naturally arise in some quantum many-body problems,
for example, in the quantum many-electron problem.

\section{Two examples of \textquotedblleft instantaneous
superluminality\textquotedblright\ in the case of electrons}

Here we illustrate the phenomenon of \textquotedblleft instantaneous
superluminality\textquotedblright\ using two examples. The first example is
the case of single electrons escaping from the interior to the exterior of a
normal metallic conductor, and the second example is the case of Cooper
pairs of electrons propagating from one end of a superconducting island to
the other.

\subsection{The deposition of charge into the interior of a normal metallic
body}

Recall Faraday's \textquotedblleft ice-pail\textquotedblright\ experiment,
which is sketched in Figure 3. Charge is being delivered into the interior
of a normal metallic body (i.e., the \textquotedblleft ice
pail\textquotedblright ) by means of a charged metal ball attached to the
end of wooden stick (i.e., an insulated rod). The charged ball is being
slowly lowered through the top opening of the ice pail until it contacts the
bottom of the pail. Upon contact, the initial charge on the ball is observed
to disappear from the ball, and also from the interior surface of the ice
pail. Furthermore, the charge from the ball is observed to \emph{suddenly}\
reappear on the exterior surface of the pail. How suddenly does this charge
transfer process occur?

\begin{figure}
\includegraphics[width=4.75in]{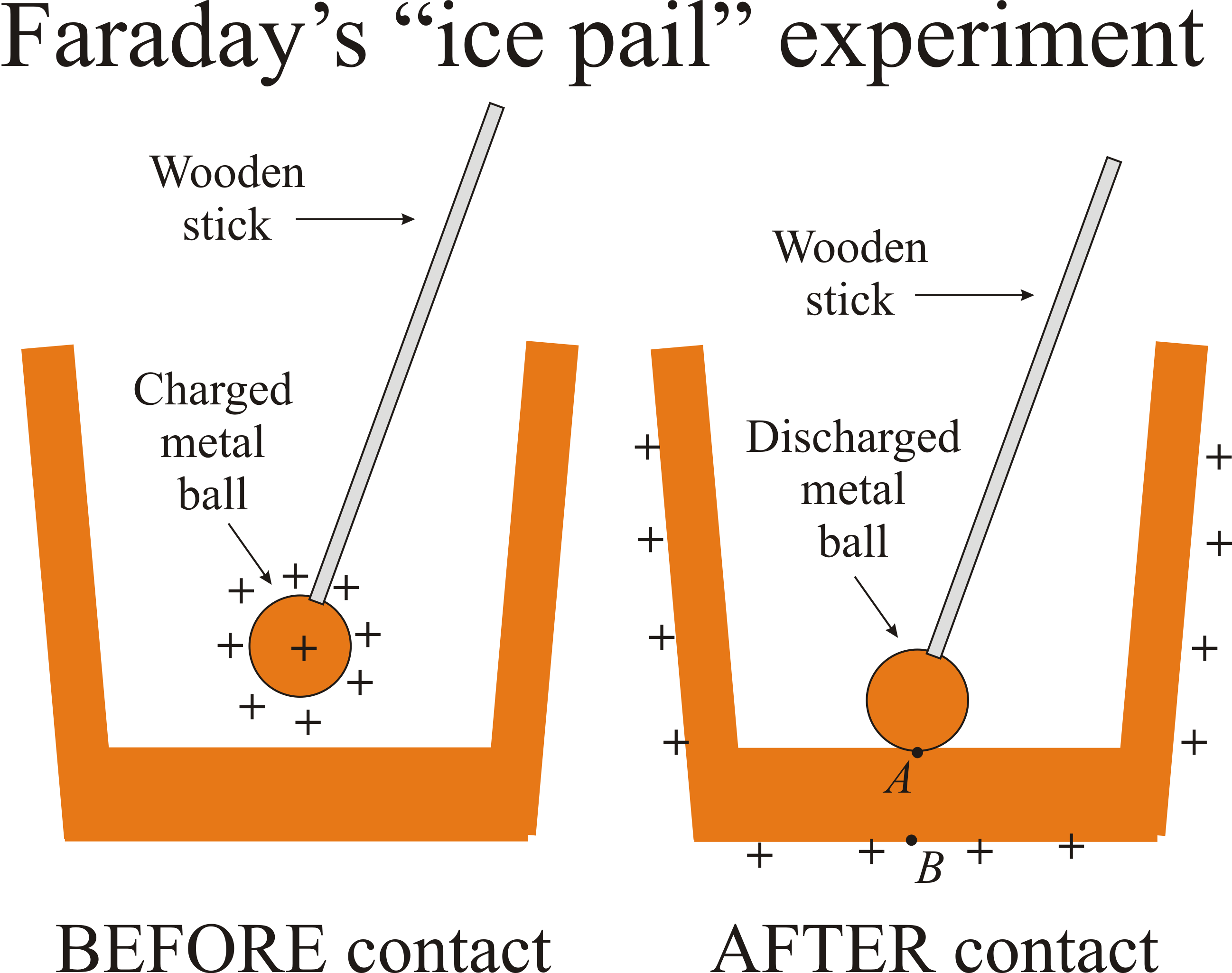}
\caption{Faraday's \textquotedblleft
ice-pail\textquotedblright\ experiment. A charged metal ball at the end of a
wooden stick is slowly and carefully lowered into the interior of a metallic
ice pail, until it contacts the bottom of the pail. Upon contact, the charge
on the ball disappears from the ball, and also from the interior of the ice
pail, and suddenly reappears on the exterior surface of the pail. Note that
the metallic bottom of the pail can be made arbitrarily thick. How suddenly
does the charge disappear from the ball and reappear on the exterior surface
of the ice pail?}
\end{figure}

There are two possible answers to this question. The first possible answer
is that the charge of the ball will initially escape from the ball along the
inside surface of the pail, and then will finally reappear on the outside
surface of the pail, after the escaping charge has propagated as \emph{%
surface} electrical currents climbing up and over the rim of the pail. Such
surface currents will propagate at the speed of light, since magnetic fields
will be generated by these currents, and therefore will cause an
electromagnetic wave to propagate along the surface of the metal. Therefore
this \emph{surface} kind of charge transfer will not be instantaneous, but
will be retarded by the speed of light.

The second possible answer is that the charge of the ball will try to escape
as a \emph{volume} electrical current directly from the point of contact of
the ball with the bottom of the pail to the nearest possible point on the
exterior of the pail (i.e., from point $A$ to point $B$ and its surroundings
in Figure 3). Note that shortest distance is that of the straight line
joining points $A$ and $B$, which lies entirely within the volume of the
metallic bottom of the pail. These \emph{volume} currents, which will flow
inside the metal of the bottom of the pail, will be driven by the electric
field lines emanating from the charge on the ball. Since no magnetic field
can be generated within the interior of any metal beyond a skin depth of the
surface of the metal, no propagating electromagnetic wave can be generated
within the volume of the metal inside the bottom of the pail. Rather, these
volume currents will be driven by the \emph{instantaneous} Coulomb electric
field lines emanating from the initial charge on the ball. Therefore this 
\emph{volume} kind of charge transfer will occur directly from $A$ to $B$.
It represents a kind of \emph{instantaneous action-at-distance}, and will
not be retarded by the speed of light.

This latter kind of \emph{instantaneous} charge transfer process is highly
counter-intuitive, and has never been observed before. Does it really exist?
In order to understand it better, consider the following \textquotedblleft
thought experiment\textquotedblright\ depicted in Figure 4, in which a
pulsed electron beam suddenly deposits charge at the center of a copper
sphere through a radial hole. A grounded cylindrical sleeve surrounding the
incoming electron beam prevents the copper sphere from seeing the
approaching electrons, until they actually strike the center of the sphere.

The continuity equation applied to the copper sphere states that%
\begin{equation}
\nabla \cdot \mathbf{j}+\frac{\partial \rho }{\partial t}=0
\label{continuity equation}
\end{equation}%
where $\mathbf{j}$ is the electrical current density flowing at any point
within the sphere and $\rho $ is the charge density at that point.

\begin{figure}
\includegraphics[width=4.75in]{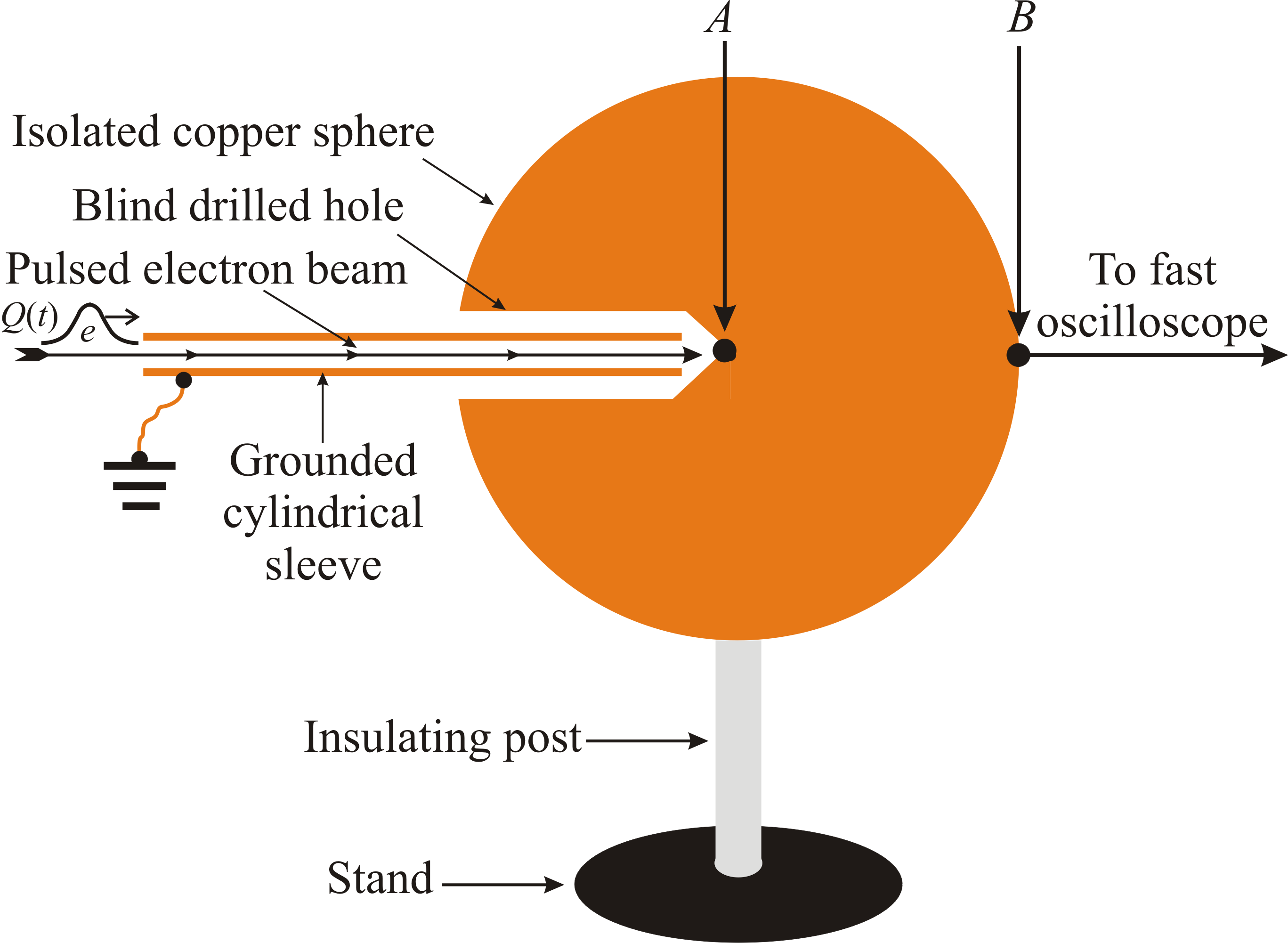}
\caption{A pulsed electron beam enters a
copper sphere (in \emph{orange}) through an insulated, grounded copper
sleeve inserted within a radial hole, and stops at the center (point $A$) of
the sphere. The grounded copper sleeve prevents the copper sphere from
seeing the incident electrons before they strike point $A$. Note that the
radius of the sphere can be made arbitrarily large. How suddenly does the
charge disappear from point $A$ and reappear at the surface, for example, at
point $B$?}
\end{figure}

Now let us assume that Ohm's law holds at every point inside the sphere, so
that%
\begin{equation}
\mathbf{j}=\sigma \mathbf{E}  \label{Ohm's law}
\end{equation}%
where $\sigma $ is the conductivity of the copper sphere, which is composed
of a homogeneous and isotropic copper material, and where $\mathbf{E}$\ is
the local electric field inside the body, which is driving the electrical
currents flowing from the center to the surface of the sphere. Actually, all
that we need to assume here is that the local current density is \emph{%
linearly related} to the local electric field that is driving the currents.

Substituting (\ref{Ohm's law}) into (\ref{continuity equation}), and using
the fact that $\sigma $ is a constant, one gets%
\begin{equation}
\nabla \cdot \left( \sigma \mathbf{E}\right) +\frac{\partial \rho }{\partial
t}=\sigma \left( \nabla \cdot \mathbf{E}\right) +\frac{\partial \rho }{%
\partial t}=\sigma \left( \frac{\rho }{\varepsilon _{0}}\right) +\frac{%
\partial \rho }{\partial t}=0
\end{equation}%
where we have used Maxwell's first equation $\nabla \cdot \mathbf{E}=\rho
/\varepsilon _{0}$. Therefore the charge density $\rho $\ obeys the linear,
first-order partial differential equation%
\begin{equation}
\frac{\partial \rho }{\partial t}=-\left( \frac{1}{\tau }\right) \rho
=-\left( \frac{\sigma }{\varepsilon _{0}}\right) \rho 
\label{PDE for decay of charge density}
\end{equation}%
which implies an exponential decay with a time constant (i.e., the
\textquotedblleft Jeans\textquotedblright\ time scale \cite{Jeans})%
\begin{equation}
\tau =\frac{\varepsilon _{0}}{\sigma }  \label{Jeans time scale}
\end{equation}%
The solution to (\ref{PDE for decay of charge density}) is the exponential
decay law%
\begin{equation}
\rho (\mathbf{r},t)=\rho (\mathbf{r},0)\exp (-t/\tau )
\end{equation}%
According to this solution, if, initially at $t=0$, the material is neutral
at any point $\mathbf{r}$ in the interior of the conducting body, i.e., if%
\begin{equation}
\rho (\mathbf{r},t=0)=0
\end{equation}%
then it follows that at the same point $\mathbf{r}$, the body must remain
neutral at all later times $t$, i.e.,%
\begin{equation}
\rho (\mathbf{r},t)=0
\label{rho is constant for all interior points}
\end{equation}%
for all $t>0$. In other words, the body at all interior points $\mathbf{r}$ within its
volume, during the entire charge transfer process from point $A$ to point $B$%
, must remain electrically neutral.

However, suppose that at a single point $A$, such as at the center in the
interior of the copper sphere in Figure 4 at time $t=0$, the conducting body
at this point were suddenly to be made non-neutral, for example, by charge
being deposited at point $A$ by a sudden charge deposition by the pulsed
electron beam (or, similarly, by a sudden contact of the charged ball with
the bottom of the ice pail at point $A$ in Figure 3), so that%
\begin{equation}
\rho (\mathbf{r}_{A},t=0)\neq 0
\end{equation}%
Then at all later times $t>0$ following this sudden charge deposition, the
charge density at this point will decay exponentially as follows:%
\begin{equation}
\rho (\mathbf{r}_{A},t)=\rho (\mathbf{r}_{A},0)\exp (-t/\tau )
\label{Solution at A}
\end{equation}%
However, all other interior points other than $A$ that were initially
electrically neutral, must remain electrically neutral at all later times $%
t>0$. This implies that currents originating from the decay of the charge at 
$A$ cannot accumulate any charge at intermediate points interior to the 
\emph{volume} of the body, since the divergence of the current density must
vanish at all interior points due to the solution (\ref{rho is constant for
all interior points}). Therefore the only points where the charge can
accumulate due to the nonvanishing current density originating from the
point charge deposition at $A$ would be at the \emph{surface} on the
exterior of the body, such as at point $B$ in Figure 4.

According to the solution (\ref{Solution at A}), the charge at point $A$
will disappear, and will extremely quickly reappear at the surface, for
example, at point $B$. Let us put in some numbers to see how quickly this
happens. For the case of copper, the measured value of the conductivity is%
\begin{equation}
\sigma _{\rm{Cu}}=59.6\times 10^{6}{\:}\rm{S}\cdot \rm{m}^{-1}
\label{conductivity of copper}
\end{equation}%
Therefore the decay time for the charge density at $A$ is predicted to occur
over the extremely short time scale \cite{limitations of Ohm's law}%
\begin{equation}
\tau =\frac{\varepsilon _{0}}{\sigma _{\rm{Cu}}}=1.48\times 10^{-19}{\:}
\rm{s}\approx 0.15{\:}\rm{ attoseconds}
\end{equation}%
This is the time that it takes light to cross a distance of%
\begin{equation}
c\tau =4.45\times 10^{-11}{\:}\rm{m}\approx 45{\:}\rm{ picometers}
\label{Bohr radius}
\end{equation}%
which is about the size of the Bohr radius (approximately 50 picometers).
Therefore for any macroscopically-sized copper sphere, the disappearance of
the electron charge at point $A$ and its sudden reappearance at an
arbitrarily far-away point $B$ in the case of an arbitrarily large sphere,
would be a clear example of superluminality.

\begin{figure}
\includegraphics[width=4.75in]{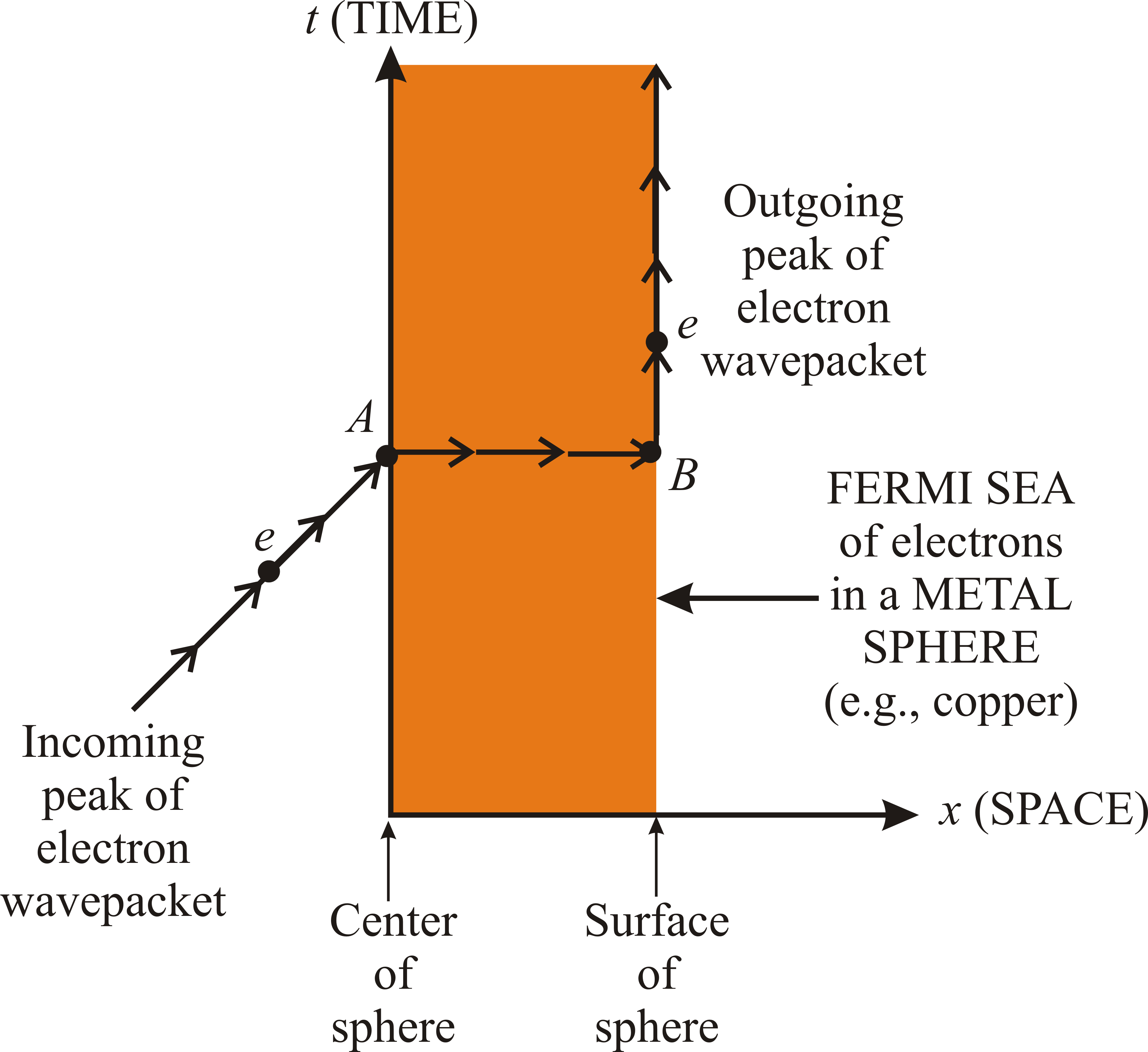}
\caption{Feynman-like space-time diagram
depicting the trajectory of the peak of a single electron wavepacket
entering the center of the metallic sphere at point $A$ in Figure 4. Here
the group velocity of the electron must be infinite between $A$ and $B$
because the electron cannot join the Fermi sea due to the Pauli exclusion
principle. It must therefore disappear at $A$\emph{,} and must
instantaneously reappear at $B$. Otherwise, charge conservation would be
violated. The horizontal trajectory between $A$ and $B$ represents a virtual
state of the electron.}
\end{figure} 

Note that no radiation can be produced in the configuration depicted in
Figure 4 due to the spherical symmetry of the time-varying currents and
charges, so that no retardation effect can occur associated with the
production of an electromagnetic wave propagating in the interior at the vacuum speed of $c$ 
\emph{within} the volume of the metal.

However, the spherical symmetry of Figure 4 is not a necessary condition for
the phenomenon of \textquotedblleft instantaneous
superluminality\textquotedblright\ to occur. In particular, the breaking of
the spherical symmetry of the conducting body by the hole drilled into the
copper sphere shown in Figure 4 for the purpose of the admission of the
electron beam, is unimportant, as evidenced by the known observations in
Faraday's \textquotedblleft ice pail\textquotedblright\ experiment depicted
in Figure 3. The configuration of Faraday's ice pail experiment clearly does
not possess the near-spherical symmetry of the configuration shown in Figure
4. In particular, note that the small, radial hole in Figure 4 can be
replaced by the very large opening at the top of the ice pail in Figure 3.
Nevertheless, the charge deposited by the metallic ball is observed to
disappear from the \emph{inside} the ice pail, and to reappear suddenly on
the \emph{outside} of the ice pail. 

If the above analysis, which is based on the continuity equation and on the
linearlity of Ohm's law, turns out to be correct, this charge transfer
process should have an unusual, instantaneously superluminal, component
arising from the internal, volume currents produced by the Coulomb field
within the metal, in addition to the usual, luminal component that one would
expect to arise from the external, surface currents associated with the
propagation of an electromagnetic wave.

At the quantum level, the above \textquotedblleft instantaneous
superluminality\textquotedblright\ effect is represented by the Feynman-like
space-time diagram shown in Figure 5, which describes the charge transfer
process for an individual electron which is initially approaching the center
of the copper sphere at point $A$ in Figure 4, with an energy less than the Fermi energy of copper. The electron will propagate superluminally via a
virtual quantum state from point $A$ to point $B$ through the copper metal,
inside which \emph{there exists a Fermi sea of identical electrons}. This
sea is in an entangled state, namely, the Slater determinant state.
Entangled states lead to nonlocal, Einstein-Podolsky-Rosen (EPR) effects, in
which instantaneous quantum correlations-at-a-distance can occur.

The electron which enters the metal at point $A$ at the center of the sphere
will be prevented by the Pauli exclusion principle from joining the Fermi
sea of identical electrons in the interior of the metal \cite{Pauli
exclusion}. Hence it has no choice but to reappear suddenly on the exterior
surface of the metal, for example, at point $B$ in Figure 4. Due to charge
conservation, the disappearance of the electron at event $A$ in Figure 5,
and therefore the disappearance of its charge $e$ at event $A$, must be 
\emph{instantaneously} accompanied the \emph{simultaneous} reappearance ---
in the reference frame of the center of mass of the Fermi sea --- of an
indistinguishable electron at event $B$, along with the reappearance of
exactly same charge $e$ at event $B$. Otherwise, charge conservation would
be violated. Note that this will be true no matter how far apart $A$ and $B$
are from each other. Hence instantaneous actions-at-a-distance, in the form
of Einstein-Podolsky-Rosen quantum correlations-at-a-distance, necessarily
follow from the conservation of charge.

However, note here that the single electron approaching the center of the
metal sphere will most probably go from the center to the surface of the
sphere, and not from the surface to the center. This is because the density
of allowed final states for the electron is much larger on
the surface of the sphere than the density of allowed initial states at the center. Note also that here the charge transfer process
is a dissipative one, and hence that it is an irreversible one.

Thus individual photons, which have earlier been observed to tunnel
superluminally through a tunnel barrier \cite{Steinberg}, are not the only
particles that can propagate through matter superluminally. Electrons can
also be transferred superluminally through matter, for example, through 
a Fermi sea.

\subsection{Experiment to observe \textquotedblleft instantaneous
superluminality\textquotedblright\ in a long aluminum bar}

We are presently performing an experiment to test the highly
counter-intuitive prediction of \textquotedblleft instantaneous
superluminality\textquotedblright\ for electron charge transfer in normal
metals. A long, thick aluminum bar (six feet long, and five inches in
diameter) has two blind holes (both about an inch deep, and half an inch in
diameter \cite{aspect ratio}) drilled into either end of the bar, so that
two miniature \textquotedblleft Faraday ice pails\textquotedblright\ can be
formed at the left and right ends of the bar, respectively (see Figure 6).
Two grounded coaxial cables are then inserted through thin, insulating
sleeves into the two small cavities thus formed at the left and right ends
of the long bar. Two small metal balls are soldered to the two ends of the
center conductors of these cables, so that electrical contact can be made at
the two points $A$ and $B$ deep inside these two miniature \textquotedblleft
Faraday ice pails\textquotedblright .

\begin{figure}
\includegraphics[width=4.75in]{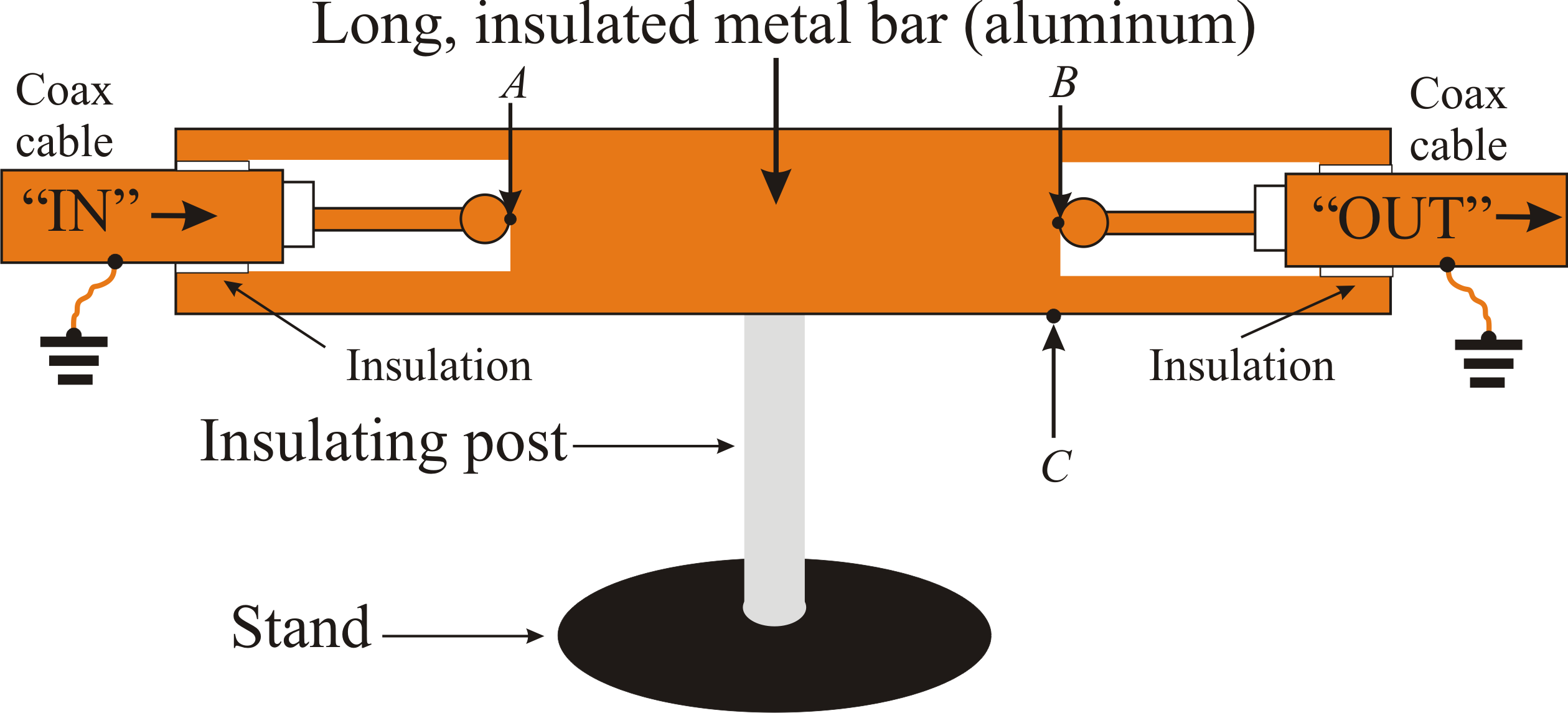}
\caption{Schematic of an experiment to
test the prediction of \textquotedblleft instantaneous
superluminality\textquotedblright\ in a long aluminum bar. The left cavity
is for the insertion of the \textquotedblleft IN\textquotedblright\ coaxial
cable in the first miniature \textquotedblleft Faraday ice
pail\textquotedblright\ on the left, and the right cavity is for the
insertion of the \textquotedblleft OUT\textquotedblright\ coaxial cable in
the second miniature \textquotedblleft Faraday ice pail\textquotedblright\
on the right. A nanosecond pulse generator is connected to the
\textquotedblleft IN\textquotedblright\ coaxial cable, and a fast
oscilloscope is connected to the \textquotedblleft OUT\textquotedblright\
coaxial\ cable. $A$ and $B$ are points of electrical contact with the center
conductors of the \textquotedblleft IN\textquotedblright\ and
\textquotedblleft OUT\textquotedblright\ coaxial cables, respectively. A
third coaxial cable (not shown) is connected to point $C$ in order to detect
the luminal signal for calibration purposes.}
\end{figure}

Charge is delivered to point $A$ by a voltage pulse which is generated by
means of a nanosecond pulse generator connected via a coaxial cable to the
\textquotedblleft IN\textquotedblright\ port on the left side of the
aluminum bar. Most of the electric field lines emanating from point $A$ that
are generated by this pulse will intersect with the outer surface of bar,
and will thus generate surface currents flowing along the exterior surface
of the bar. Since time-varying magnetic fields will be generated by these
surface currents, one expects an electromagnetic wave to propagate along the
surface from the left to the right end of the bar, in the usual TEM\ mode of
propagation. Thus a luminal pulse traveling at the vacuum speed of light
should result from such surface currents.

However, some of the electric field lines emanating from point $A$ will take
the shortest possible path to reach point $B$, including the straight-line
path that directly joins $A$ to $B$. Such electric field lines will stay
within the interior volume of the metal rod, and drive an internal current
density through the ohmic relationship $\mathbf{j}=\sigma \mathbf{E}$, where 
$\sigma $ is the conductivity of aluminum. The analysis starting from the
continuity equation given from (\ref{continuity equation}) to (\ref{Bohr
radius}) leads to the conclusion that the extremely short Jeans time scale (%
\ref{Jeans time scale}) should hold for these volume currents.

One expects that a small fraction of the total number of electric field
lines emanating from the charge deposited at point $A$ will remain deep
inside the volume of the metal rod at interior points during their journey
from $A$ to $B$. This includes the straight electric-field line that
directly connects $A$ to $B$. The fraction of internal electric field lines
will be approximately given by the ratio of the solid angle subtended by the
midsection of rod with respect to source point $A$, to $4\pi $ steradians. 

Therefore for the dimensions of our aluminum bar, we expect an attenuation
of the voltage amplitude of about a factor of a thousand in the transmission
of the instantaneously superluminal signal from $A$ to $B$, relative to the
voltage amplitude of the luminal signal, which can be picked up at point $C$%
. However, the resulting signal-to-noise ratio for detecting the pulse at $B$
by means of a fast oscilloscope connected to the \textquotedblleft
OUT\textquotedblright\ coaxial cable, should still be large enough to allow
for a significant detection of the instantaneously superluminal signal. For
the purposes of calibration, a third coaxial cable will be connected to
point $C$ of Figure 6 on the surface of the rod near point $B$, so that the
luminal surface-current signal can also be detected and displayed on
separate channel of the fast oscilloscope for a direct comparison with the
superluminal signal. 

The data and the data analysis of this experiment will be presented
elsewhere.

\subsection{Superluminal charge transfer of Cooper pairs through a
superconducting island}

Figure 7 illustrates a second example of \textquotedblleft instantaneous
superluminality\textquotedblright\ for electrons. Consider a superconducting
circuit consisting of a long superconducting island, which is connected to a
charge source by means of a Josephson tunnel junction at point $A$ on its
left end, and to charge measuring device by means of an identical Josephson
tunnel junction at point $B$ on its right end.

\begin{figure}
\includegraphics[width=4.75in]{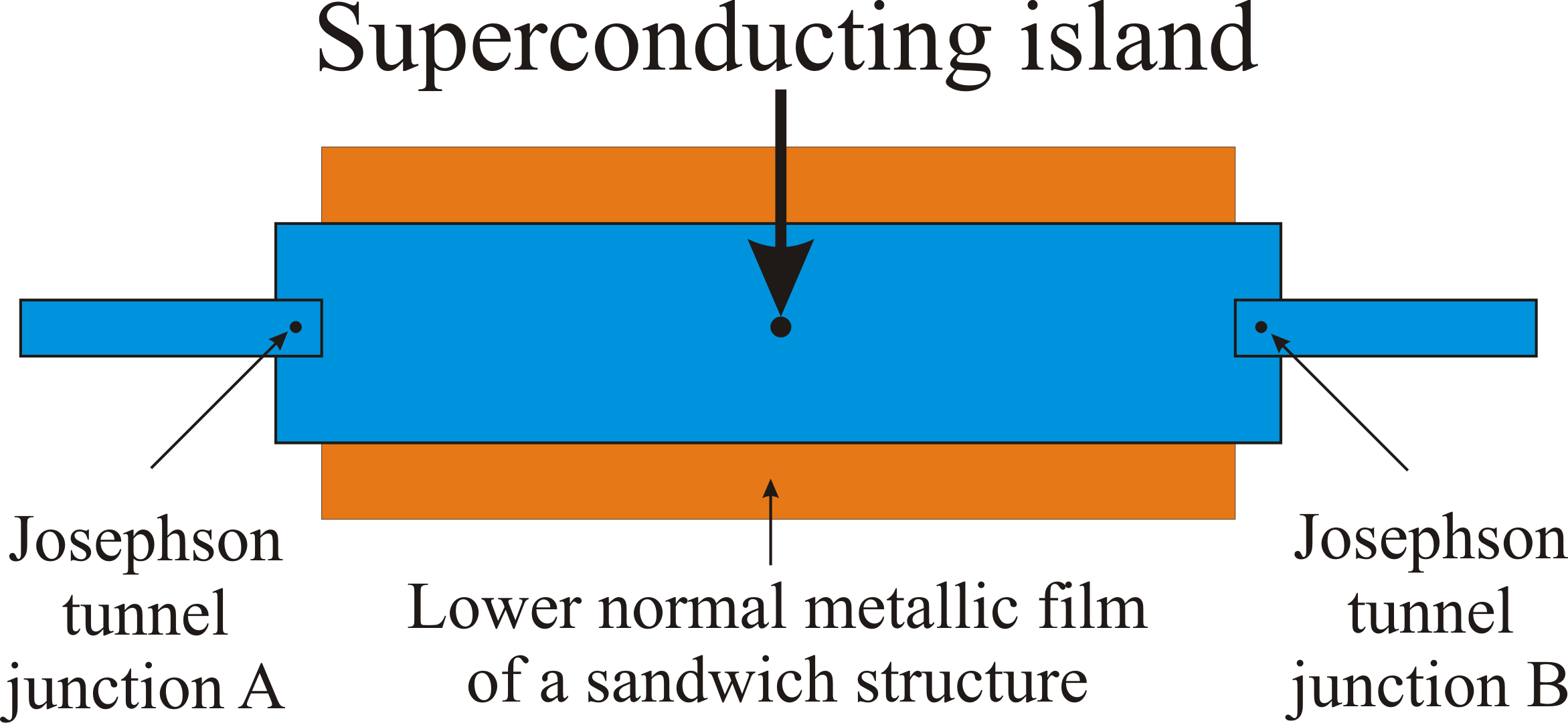}
\caption{Charge is being transferred from
point $A$ to point $B$ of a long superconducting island (in \emph{blue}). A
lower normal metallic film (in \emph{orange}) lies in intimate electrical
contact with it just beneath the island. Not shown is an upper normal
metallic film lying just above the island in intimate contact with it, in a
bimetallic \textquotedblleft sandwich\textquotedblright\ structure. These
normal films greatly diminishes the capacitance of the island, so that
\textquotedblleft Coulomb blockade\textquotedblright\ occurs.}
\end{figure}

The superconducting island is sandwiched tightly between upper and lower
normal metallic films (made out of copper, for example), which are therefore
in intimate electrical contact with it in a bimetallic structure. These upper and lower normal films
serve to greatly decrease the island's capacitance $C_{\rm{island}}$ \cite%
{image charge}. 

The charging energy for depositing even just a single charge of $Q_{\rm{%
Cooper{\,}pair}}=2e$ (i.e., the charge a single Cooper pair) onto the
superconducting island is given by%
\begin{equation}
U_{\rm{charge}}=\frac{1}{2}\frac{Q_{\rm{Cooper{\,}pair}}^{2}}{C_{\rm{%
island}}}
\end{equation}%
Since the capacitance of the island can be made very low, the charging
energy $U_{\rm{charge}}$ for adding even just a single Cooper pair to the
island can be made very large when compared to the Josephson junction
coupling energy%
\begin{equation}
U_{\rm{Josephson}}=I_{\rm{J}}\Phi _{0}
\end{equation}%
where $I_{\rm{J}}\,$is the critical current of the Josephson junctions and 
$\Phi _{0}=h/2e$ is the quantum of flux. Therefore it becomes highly energetically
unfavorable for the superconducting island to become charged even by a
single Cooper pair. Hence, due to the \emph{discreteness} of electrical
charge that arises from the quantization of charge, the island will remain
electrically neutral at all times. This leads to a phenomenon called
\textquotedblleft Coulomb blockade\textquotedblright , in which all Cooper
pairs are effectively \textquotedblleft blockaded\textquotedblright\ from
entering the island. Note that this effect is independent of the distance
separating $B$ from $A$.

\begin{figure}
\includegraphics[width=4.75in]{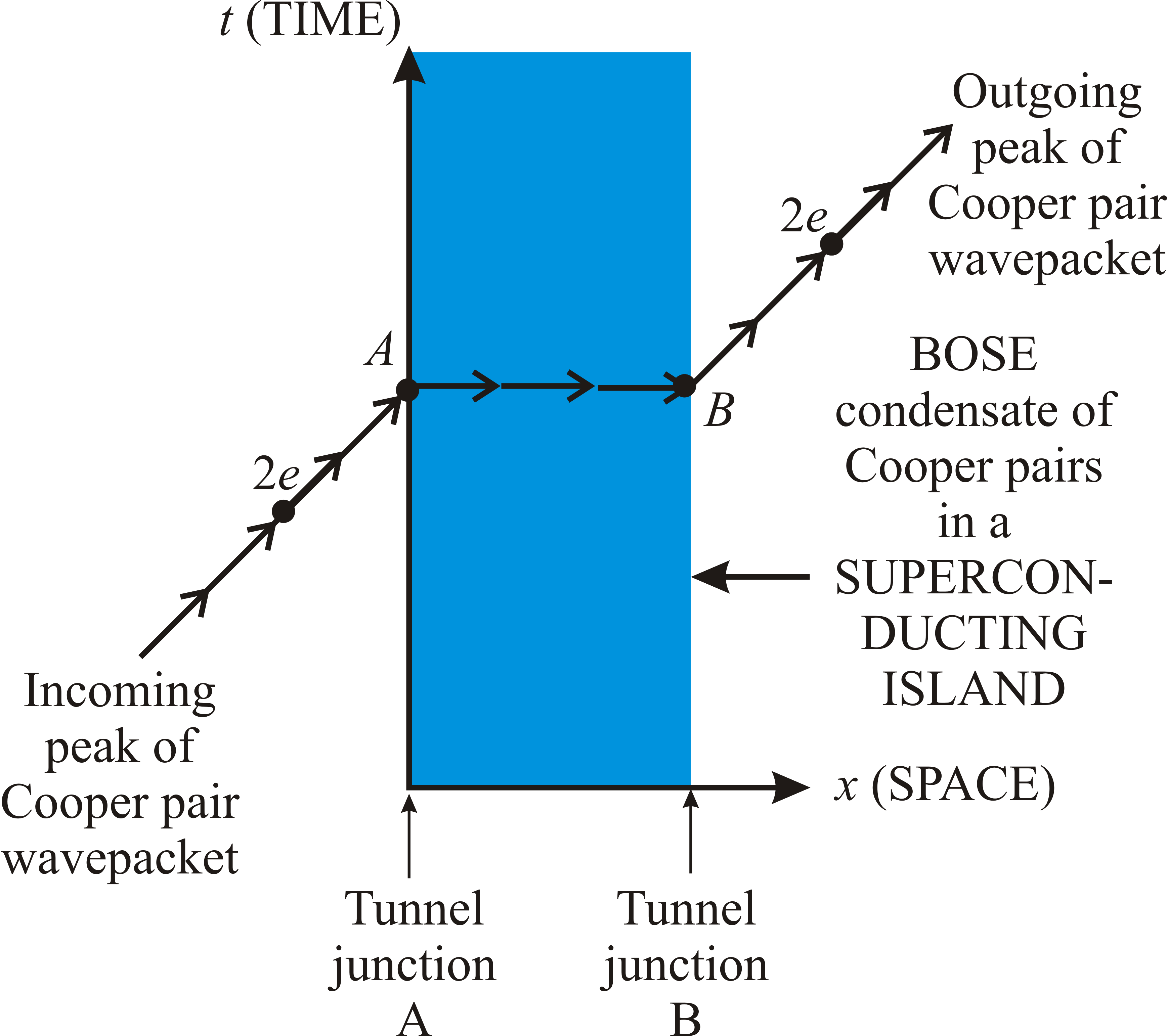}
\caption{Feynman-like space-time diagram of the \textquotedblleft instantaneously
superluminal\textquotedblright\ charge transfer process of a Cooper pair
through the superconducting island of Figure 7 from tunnel junction $A$ to
tunnel junction $B$. The horizontal trajectory between $A$ and $B$ represents a virtual state of the Cooper pair. This is another example of a quantum-mechanical
\textquotedblleft instantaneous action-at-a-distance\textquotedblright .}
\end{figure}

As a consequence of the \textquotedblleft Coulomb blockade\textquotedblright  effect, a
Cooper pair entering from the left at tunnel junction $A$ will have to
instantaneously exit to the right through tunnel junction $B$ so as to maintain
the charge neutrality of the superconducting island, no matter how far $B$
is from $A$. This kind of instantaneous charge transfer from $A$ to $B$
leads to yet another example of a Feynman-like space-time diagram with the
property of \textquotedblleft instantaneous
superluminality\textquotedblright , which is\ sketched in Figure 8. This
diagram represents the instantaneous charge transfer of \emph{Cooper pairs}
of electrons through a Bose condensate of these pairs, a process that leaves
the net charge of the condensate electrically neutral at all times (remember that the Cooper
pairs initially present on the island are neutralized by the background
ionic lattice).

An important difference between \textquotedblleft instantaneous
superluminality\textquotedblright\ in the case of superconductors, as
compared to the case of normal conductors, is the fact that in former, the
phenomenon is a non-dissipative one, whereas in the latter, it is a
dissipative one (compare Figures 8 and 5). However, since an experiment can
be performed at room temperature in normal metals, \textquotedblleft
instantaneous superluminality\textquotedblright\ will be much easier to
demonstrate in the former case than in the latter case. Nevertheless,
experiments in the superconducting case could yield larger superluminal
signals, since the dilution factor arising from the solid-angle
considerations of the normal metal case would not apply.

\section{Sommerfeld's front velocity and causality in special relativity}

In light of the predictions of the above superluminal phenomena in quantum
many-electron systems, the question naturally arises: Do they violate
relativity? In order to answer this question, we need to introduce the
important concept of the \textquotedblleft front velocity\textquotedblright
, which is due to Sommerfeld \cite{Brillouin}.

Consider an electromagnetic carrier wave of the form%
\begin{equation}
E(z,t)=E_{0}\cos (kz-\omega t)\Theta (t-z/c)
\label{E field times theta function}
\end{equation}%
\begin{equation}
B(z,t)=B_{0}\cos (kz-\omega t)\Theta (t-z/c)
\label{B field times theta function}
\end{equation}%
where the theta function is defined as follows: $\Theta (t^{\prime })=0$ for all times $t^{\prime }<0$ and $\Theta (t^{\prime })=1$ for all times 
$t^{\prime }\geq 0$. The instant $t=0$ corresponds to the sudden turn-on of a carrier wave,
initiated, for example, by the pushing of the \textquotedblleft
ON\textquotedblright\ button of a continuous-wave signal generator located
at $z=0$. Thus one could characterize the electromagnetic wave of the form
given by (\ref{E field times theta function}) and (\ref{B field times theta
function}) as the \textquotedblleft push-the-button\textquotedblright\
signal waveform, and one could think of the \textquotedblleft front
velocity\textquotedblright\ as the \textquotedblleft
push-the-button\textquotedblright\ velocity. This discontinuous kind of
waveform can subsequently enter into any kind of medium, but the
discontinuity of the theta function, that is, the the wave front associated
with the original sudden turn-on of the carrier wave, will always travel
within the medium at the vacuum speed of light.

The discontinuity represented by the theta function in (\ref{E field times
theta function}) and (\ref{B field times theta function}) contains Fourier
components at infinite frequency, or equivalently, at infinitely high
energy. However the index of refraction of electromagnetic waves at infinite
frequencies in all types of media will universally approach unity, i.e.,%
\begin{equation}
n(\omega \rightarrow \infty )\rightarrow 1
\end{equation}%
independent of the medium, since any medium will behave exactly like the vacuum when it is excited by
electromagnetic waves at infinite frequencies. In other words, since we know
that the speed of electromagnetic waves in the vacuum is exactly $c$, it
follows that the phase velocity at infinite frequencies, which is equivalent
to that of the front velocity, must universally also be exactly the vacuum
speed of light $c$, independent of the nature of the medium.

In this way, Sommerfeld showed that it is the front velocity, and only the
front velocity, that relates a cause to its effect in special relativity.
The theta function in (\ref{E field times theta function}) and (\ref{B field
times theta function}) is what guarantees that no effect can precede its
cause. The light-cone structure of spacetime in relativity follows from the
propagation of \textquotedblleft signals\textquotedblright\ at the front
velocity, and not from the propagation of \textquotedblleft
signals\textquotedblright\ at the group velocity. Hence the
\textquotedblleft signal\textquotedblright\ velocity of physics, in the
fundamental sense of a \textquotedblleft signal\textquotedblright\ that
connects a cause to its effect, is given by the front velocity, and not by
the group velocity.

In the two specific examples of \textquotedblleft instantaneous
superluminality\textquotedblright\ in the case of electrons, one for normal
metals and the other for superconductors, one must again ask: Do they
violate relativistic causality? The answer is again no, because the front
velocity in these two examples will be given by the velocity for electrons
with infinitely high energies, i.e., with energies much larger than the
Fermi energy of the normal metal, or of the BCS gap energy of
superconductors. Such high energy electrons will pass through these metals with a front velocity equal to the vacuum speed of light.
Again, relativistic causality is related only to \textquotedblleft
signals\textquotedblright\ which can be transported by these extremely
high-energy electrons, and not by the infinite group velocities of the
low-energy electrons depicted in the Feynman-like diagrams of Figures 5 and
8.

\textbf{Acknowledgments:} I thank Luis Martinez, Steve Minter, Robert Haun,
and Kirk Wegter-McNelly for their help.


\begin{thebibliography}{99}
\bibitem{Pauli exclusion} The Pauli exclusion principle applies in
configuration space as well as in momentum space. The Fourier transform of
the Fermi sea in momentum space leads to a Fermi sea in configuration space
with a uniform density of electrons with a uniform negative charge density
in the interior of the metal that compensates for the uniform positive
charge density of the ionic lattice, except near the surface of the metal on
the length scale $1/k_{F}$ where $k_{F}$ is the Fermi wavenumber. The
incoming electron is excluded by the Pauli exclusion principle from the
joining the Fermi sea in momentum space. Hence it will also be excluded from
joining the Fermi sea in configuration space, except near the surface of the
metal on a length scale of $1/k_{F}$.
\bibitem{aspect ratio} Note that the aspect ratio of the blind holes drilled
into the aluminum bar (i.e., the ratio of the depth of the hole to its
diameter) is around 2:1, which is comparable to the aspect ratio of the
\textquotedblleft ice pail\textquotedblright\ used by Faraday in his
original experiments.
\bibitem{image charge} A single Cooper pair will be attracted to its own
image charge, which will be formed inside the nearby normal metal film in
the bimetallic sandwich structure of Figure 7. However, the BCS energy gap will prevent
the breakup of the Cooper pair upon its attempt to cross over into the
normal metal film in order to reach the vacuum side of the film, and thereby
to charge up the capacitor formed by the superconducting island structure.
Hence the effective capacitance of the bimetallic sandwich structure shown
in Figure 7 will be much lower than that for a configuration without the
bimetallic sandwich structure. The suppressed capacitance of the island
structure should be on the order of $C_{\rm{island}}\simeq \varepsilon
_{0}\xi \simeq 10^{-19}$ f, where $\xi \simeq 10$ nm is the coherence length
of the superconductor.
\end{thebibliography}
\end{document}